\documentclass[iop]{emulateapj}

\usepackage{color, colortbl}
\usepackage{rotating}
\usepackage{txfonts}
\usepackage{graphicx}
\usepackage{subfigure}
\usepackage{multirow}
\usepackage{verbatim}
\usepackage{longtable}
\usepackage{url}
\usepackage{float}
\usepackage{ esint }
\usepackage{dcolumn}
\usepackage{hyperref}
\usepackage[utf8]{inputenc}
\makeatletter
\newcommand\footnoteref[1]{\protected@xdef\@thefnmark{\ref{#1}}\@footnotemark}
\makeatother

\slugcomment{Submission draft}

\shorttitle{Robo-AO \textit{Kepler} Planetary Candidate Survey V}
\shortauthors{Ziegler et al.}

\begin{document}

\title{Robo-AO Kepler Survey V: The effect of physically associated stellar companions on planetary systems}

\author{Carl Ziegler\altaffilmark{1}, Nicholas M. Law\altaffilmark{1}, Christoph Baranec\altaffilmark{2}, Ward Howard\altaffilmark{1}, Tim Morton\altaffilmark{3}, Reed Riddle\altaffilmark{4}, Dmitry A. Duev\altaffilmark{4}, Ma\"{ı}ssa Salama\altaffilmark{2}, Rebecca Jensen-Clem\altaffilmark{5}, S. R.
Kulkarni\altaffilmark{4}}

\email{carlziegler@unc.edu}
\altaffiltext{1}{Department of Physics and Astronomy, University of North Carolina at Chapel Hill, Chapel Hill, NC 27599-3255, USA}
\altaffiltext{2}{Institute for Astronomy, University of Hawai`i at M\={a}noa, Hilo, HI 96720-2700, USA}
\altaffiltext{3}{Department of Astrophysical Sciences, Princeton University, Princeton, NJ 08544, USA}
\altaffiltext{4}{Division of Physics, Mathematics, and Astronomy, California Institute of Technology, Pasadena, CA 91125, USA}
\altaffiltext{5}{University of California, Berkeley, 510 Campbell Hall, Astronomy Department, Berkeley, CA 94720, USA}

\begin{abstract}
The \textit{Kepler} light curves used to detect thousands of planetary candidates are susceptible to dilution due to blending with previously unknown nearby stars. With the automated laser adaptive optics instrument, Robo-AO, we have observed 620 nearby stars around 3857 planetary candidates host stars. Many of the nearby stars, however, are not bound to the KOI. In this paper, we quantify the association probability between each KOI and detected nearby stars through several methods. Galactic stellar models and the observed stellar density are used to estimate the number and properties of unbound stars. We estimate the spectral type and distance to 145 KOIs with nearby stars using multi-band observations from Robo-AO and Keck-AO. We find most nearby stars within 1\arcsec ~of a \textit{Kepler} planetary candidate are likely bound, in agreement with past studies. We use likely bound stars as well as the precise stellar parameters from the California \textit{Kepler} Survey to search for correlations between stellar binarity and planetary properties. No significant difference between the binarity fraction of single and multiple planet systems is found, and planet hosting stars follow similar binarity trends as field stars, many of which likely host their own non-aligned planets. We find that hot Jupiters are $\sim$4$\times$ more likely than other planets to reside in a binary star system. We correct the radius estimates of the planet candidates in characterized systems and find that for likely bound systems, the estimated planetary candidate radii will increase on average by a factor of 1.77, if either star is equally likely to host the planet. We find that the planetary radius gap is robust to the impact of dilution, and find an intriguing 95\%-confidence discrepancy between the radius distribution of small planets in single and binary systems.

\end{abstract}


\section{Introduction}

The \textit{Kepler} telescope detected over 4000 planetary candidates during its four-year primary mission \citep{borucki10, borucki11a, borucki11b, batalha13, burke14, rowe14, coughlin15, dr25}. Each of these planet candidates (\textit{Kepler} objects of interest, or KOIs) requires further ground-based, high-angular resolution observations to detect contaminating nearby stars that lie within the same photometric aperture as the planet candidate host star. The additional flux from these stars dilute the \textit{Kepler} light curves and can lead to inaccurate host star characterization \citep{dressing13, santerne13} and underestimated planetary radii \citep{morton11}. Therefore, every \textit{Kepler} planetary candidate must be validated with ground-based high-angular resolution observations.

A single comprehensive survey of every KOI on a large aperture telescope would necessitate an enormous time allocation due to the inefficiencies of a conventional high-resolution instrument. Nevertheless, the community has contributed considerable effort to perform these vital observations \citep{howell11, adams12, adams13, lillo12, lillo14, horch12, horch14, marcy14, dressing14, gilliland15, wang15a, wang15b, torres15, everett15, kraus16, furlan16}. While many of these surveys were performed on large-aperture telescopes, sensitive to close (tens of mas separation) and faint (8-10 magnitudes fainter than the host star) nearby stars, any individual survey observed only a fraction of the total number of KOIs. In addition, the piecemeal approach resulted in redundant observations of a small set of KOIs (often the brightest targets). In total, less than half of the KOIs have been observed in high-resolution within these surveys.

Robo-AO, the first fully automated laser adaptive optics system, achieves an order-of-magnitude increase in time-efficiency compared to conventional high-resolution instruments. We are using Robo-AO to perform high-resolution imaging of every KOI system, sensitive to nearby stars at separations as close as 0\farcs15 and up to 6 mag fainter than the host star. The survey, covered in \citet[hereafter Paper I]{law14},  \citet[hereafter Paper II]{baranec16}, \citet[hereafter Paper III]{ziegler16}, and \citet[hereafter Paper IV]{ziegler18}, has observed 3857 KOIs to date, approximately 95\% of the planetary candidates discovered with \textit{Kepler}, and detected 620 nearby stars.

Previous studies suggest the presence of a stellar companion these systems will shape the properties of the planetary candidates \citep[e.g.,][]{katz11, naox12, wang14, kraus16}. For instance, we found low-significance evidence in Paper III that hot Jupiters are more likely to be found around KOIs with a nearby star, and that single and multiple transiting planet systems have similar nearby star fraction rates. Simulations \citep{horch14} and observations \citep{hirsch17, atkinson16} of \textit{Kepler} multiple star systems suggest, however, that a significant fraction of stars detected with Robo-AO at wide separations are unbound asterisms. The diluting effect of these unassociated nearby stars in our sample limit the ability to measure correlations between stellar binarity and the properties of the planetary systems. In this paper, we describe our analysis into the probability of association of individual systems with additional Robo-AO observations and survey simulations. The likely bound systems which host planet candidates are then used to search for insight into the effect binary stars have on the formation and evolution of planetary systems.

We begin in Section \ref{sec:targetselection} by briefly describing the Robo-AO system and the Robo-AO and Keck-AO observations. We discuss the impact that stellar companions can have on the estimated planetary radii in Section \ref{sec:datareduction}. In Section \ref{sec:stellarcharazterization}, we describe the analysis used to estimate the spectral type and distances to the KOIs and their companions, then quantify the probability of physical association of multiple systems. We discuss the results of the nearby star characterization and the implications of binary systems on the \textit{Kepler} planetary candidates in Section \ref{sec:Discussion}.  We conclude in Section \ref{sec:conclusion}.

\section{Survey Observations and Analysis}
\label{sec:targetselection}

We use Robo-AO and Keck-AO to observe KOIs with detected nearby stars in multiple bands to estimate the spectral type and distance to each star. These properties allow us to quantify the probability of association between the primary and secondary components in each system.

\subsection{Robo-AO}\label{sec:observations}
Observations in the survey were performed using the Robo-AO automated laser adaptive optics system at Palomar and Kitt Peak \citep{baranec14, baranec17, jc18} that can efficiently perform large, high angular resolution surveys. The AO system runs at a loop rate of 1.2 kHz to correct high-order wavefront aberrations, delivering median Strehl ratios of 9\% and 4\% in the \textit{i}\textsuperscript{$\prime$}-band at Palomar and Kitt Peak, respectively. Observations were taken in a long-pass filter cutting on at 600 nm (LP600 hereafter). The LP600 filter approximates the \textit{Kepler} passband at redder wavelengths, while also suppressing blue wavelengths that reduce adaptive optics performance. The LP600 passband is compared to the \textit{Kepler} passband in Figure 1 of Paper I. We obtained high-angular-resolution images of 3313 KOIs with Robo-AO between 2012 July 16 and 2015 June 12 (UT) at the Palomar 1.5m telescope. We observed 532 additional KOIs with Robo-AO between 2016 June 8 and 2016 July 15 (UT) at the Kitt Peak 2.1m telescope.

Further follow-up observations in \textit{r}\textsuperscript{$\prime$}, \textit{i}\textsuperscript{$\prime$}, and \textit{z}\textsuperscript{$\prime$} bands of 145 KOIs with nearby stars detected by Robo-AO in previous papers in the survey were performed between 2017 March 16 and 2017 June 08 (UT) at Kitt Peak by Robo-AO.  These observations facilitate characterization of the nearby stars, and allow estimation of the association probability between the KOI and nearby star, described in Section$~\ref{sec:probassociation}$. Photometry from these observations is available in Table$~\ref{tab:distances}$.
\begin{table}
\renewcommand{\arraystretch}{1.3}
\begin{longtable}{ll}
\caption{\label{tab:survey_specs}The specifications of the Robo-AO KOI survey}
\\
\hline
KOI targets    	& 3857 \\
FWHM resolution   	& $\sim$0$\farcs$15 (@600-750 nm) \\
Observation wavelengths & 600-950 nm\\
Detector format & 1024$^2$ pixels\\
Pixel scale & 43 mas/pix (Palomar)\\
& 35 mas/px (Kitt Peak)\\
Exposure time & 90 seconds \\
Targets observed / hour & 20\\
Observation dates & 2012 July 16 --\\at Palomar 1.5m &  2015 June 12\\

Observation dates & 2016 June 8 --\\at Kitt Peak 2.1m &  2016 July 15\\
\hline
\label{tab:specs}
\end{longtable}
\end{table}

\subsection{Keck LGS-AO}
\label{sec:keckao}

We observed 10 candidate multiple systems in J, H, and K with the NIRC2 camera behind the Keck-II laser guide star adaptive optics system \citep{KeckLGS1, KeckLGS2}, on 2017 Aug 8-10 (UT). These 10 multiple KOI systems were targeted due to uncertainty in association between the primary and secondary stars with only Robo-AO visible-band photometry. Typically, three 30 s exposures were taken in each band, for a total exposure time of 270 s. The images were corrected for geometric distortion using the NIRC2 distortion solution of \citet{Yelda10}. The additional NIR photometry for multi-band observations with Keck are detailed in Table$~\ref{tab:distances}$. 



\subsection{Data Reduction and Analysis}
\label{sec:datareduction}
With the largest adaptive optics dataset yet assembled by Robo-AO, the data reduction process was automated as much as possible for efficiency and consistency. As described in Paper IV, after initial pipeline reductions, the properties of the detected nearby stars were measured. Systems that have multi-band observations with Robo-AO are characterized as described in Section \ref{sec:stellarcharazterization} to quantify the probability of association and correct the estimated planetary radius.

\subsection{Updated Transiting Object Parameters}
\label{sec:updated_radii}

A nearby star in the same photometric aperture as the target star will dilute the observed transit depth, resulting in underestimated radius estimates. We re-derive the estimated planetary radius around the 145 systems re-observed with Kitt Peak for two scenarios: the planet orbits the target star or the planet orbits the secondary star, whether unbound or bound to the primary.  For the first case, we use the relation from Paper I to correct for the transit dilution,
\begin{equation}
R_{p,A}=R_{p,0}\sqrt{\frac{1}{F_{A}}}
\end{equation}
where R$_{p,A}$ is the corrected radius of the planet bound around the primary star, R$_{p,0}$ is the original planetary radius estimate based on the diluted transit signal, and F$_{A}$ is the fraction of flux within the aperture from the primary star.  For the case where the planet candidate is bound to the secondary star, we use the relation
\begin{equation}
R_{p,B}=R_{p,0}\frac{R_{B}}{R_{A}}\sqrt{\frac{1}{F_{B}}}
\end{equation}
where R$_{B}$ and R$_{A}$ are the stellar radii of the secondary and primary star, respectively.  

The fluxes of all observed sources within the \textit{Kepler} aperture were summed to estimate the transit dilution.  We derive stellar radius estimates with relations from \citet{habets81} using spectral types from the stellar characterization described in Section \ref{sec:probassociation}.  Revised planetary radius estimates are detailed in Table$~\ref{tab:radii}$.

\section{Measuring the Probability of Physical Association}
\label{sec:stellarcharazterization}

In this section, we use several methods to determine the association probability for multiple KOI systems. We first determine the spectral type and distance to stars in each system using multi-band photometry. We then use galactic stellar models to determine the properties of likely bound and unbound stars and estimate the number of unbound stars within our sample using the stellar density observed in Robo-AO images. Lastly, we use this analysis to correct the estimated radii of planetary candidates within these systems.

\subsection{Photometric-Distance-Based Association}
\label{sec:probassociation}
Multi-band observations with an adaptive optics instrument can allow characterization of the stars detected near KOIs, giving estimates of the stars intrinsic brightness and approximate distances. If the distance estimates between the primary and a nearby star are in agreement, it is highly probable the two are in fact gravitationally bound.

We characterize the stars nearby 145 KOIs re-observed with Robo-AO, described in Section$~\ref{sec:observations}$. We targeted stars with surface gravities consistent with dwarf stars: log$~g>$3 and log$~g<$5, as estimated by \citet{dr25}.  In addition to visible photometry from Robo-AO images, we used PANSTARRs \textit{g}\textsuperscript{$\prime$}-band photometry \citep{panstarrs} for widely separated stars (typically $\rho>3\arcsec$) and extant NIR photometry from previous seeing limited and high-resolution surveys \citep{atkinson16, kraus16, furlan16}. Photometry of the blended systems was obtained from the stellar properties described in \citet{dr25}.

To estimate the spectral types of the KOIs and nearby stars, we follow the analysis described in \citet{atkinson16}. A Gaussian distribution for each available photometric color is generated based on the measured or published errors. Distributions are corrected for extinction using the standard relations from \citet{cardelli89}. These distributions are then fit to SED models \citep{kraus07}, originally assembled from a heterogeneous set of models and data for an investigation of the Praesepe and Coma Berenices, to determine spectral type.  We assume that all nearby stars lie on the main-sequence; we discuss possible background giant star contamination in Section$~\ref{sec:giantcontamination}$. For each star, we use the intrinsic brightness of the estimated spectral type in each band compared to the observed apparent magnitudes of the star to estimate the distance to that star. The average of the estimated distances from all observed bands provides the final distance estimate. Distance uncertainties are derived from repeating the spectral fits and distance estimations using photometry in each band drawn at random from the respective Gaussian distribution. The final uncertainty is the standard deviation of the resulting distribution of distance estimates from 10,000 such fits.

\begin{figure}
\includegraphics[width=0.4
\paperwidth]{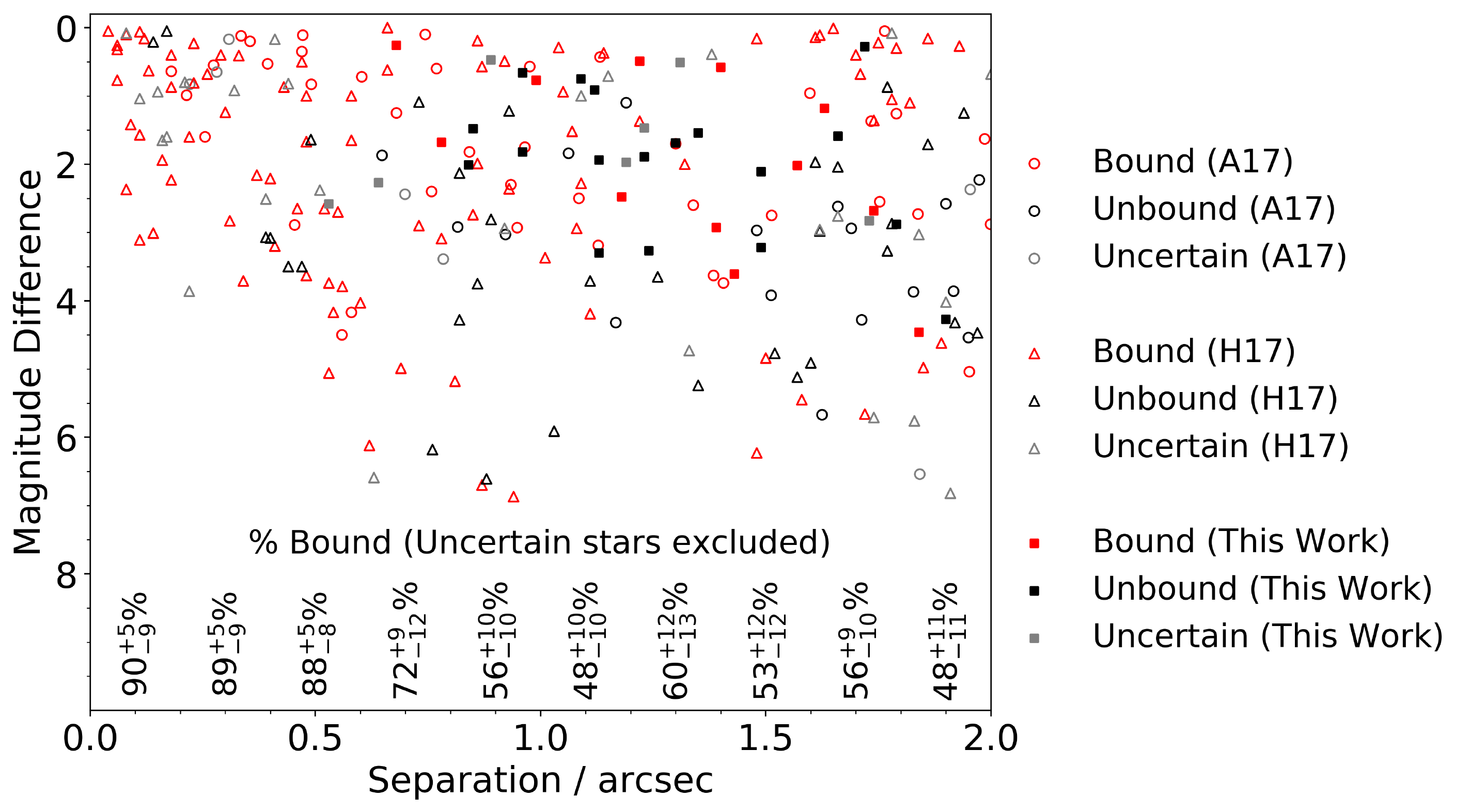}
\caption{Results of association analyses of nearby stars to KOIs from \citet[A17]{atkinson16}, \citet[H17]{hirsch17}, and this work, described in Section$~\ref{sec:probassociation}$.  The percent of nearby stars that are bound in each 0\farcs2 bin is displayed along the bottom.  Most stars within 1\arcsec of the KOI are bound, with wider separated stars more likely to be unbound.}
\label{fig:observed_systems}
\end{figure}

The resulting best-fit spectral type and distance estimates, along with measured photometry, for the 145 KOIs and nearby stars are detailed in Table$~\ref{tab:distances}$.  We combine these results with those of \citet{atkinson16} and \citet{hirsch17} to estimate the percent of nearby stars that are bound, displayed in Figure$~\ref{fig:observed_systems}$.  For results from this work and from \citet{atkinson16}, bound systems have uncertainties between the estimated distance of the primary and secondary star less than 2$\sigma$, uncertain have uncertainties between 2 and 3$\sigma$, and unbound have uncertainties greater than 3$\sigma$.  The combined sample supports the conclusion of \citet{hirsch17} and this work in Section$~\ref{sec:trilegal}$ that most stars within 1\arcsec of the primary star are bound, with the percent of stars bound decreasing at wider separations.

\subsection{Association Probabilities from Simulations}
\label{sec:trilegal}

We can estimate the probability of association of a nearby star based on its separation and magnitude difference to the primary star using the TRILEGAL Galactic stellar model \citep{trilegal}. Following a similar analysis to \citet{horch14}, we simulate star fields for ten one-square-degree star fields randomly distributed in the \textit{Kepler} field of view.  To match the distribution of stellar characteristics of the KOIs, we limit our sample to distances within 1300 pc, stellar effective temperatures between 3,000 and 10,000 K, and surface gravity (log$~g$) between 3.3 and 4.7.  The majority of KOIs are solar-type stars \citep{batalha10}, thus binaries were populated at a companion rate of 46$\%$, a fraction determined from observations for solar-types stars by \citet{duquennoy91} and \citet{raghavan10}.  Orbital periods of the companion stars were drawn at random from the log-normal distribution from \citet{duquennoy91}.  Eccentricities were also drawn from the distribution found in \citet{duquennoy91}.  The semi-major axis of the orbit was determined from the stellar masses and period, and we select random values for the cosine of inclination (cos$i$), ascending node ($\Omega$),  the angle in the orbit between the line of nodes and the semi-major axis ($\omega$), and the time of periastron passage.  The companion stars are than placed at an angular distance from the primary using by converting the true orbital distance and the distance from the solar system to the stars.

\begin{figure}
\includegraphics[width=0.4
\paperwidth]{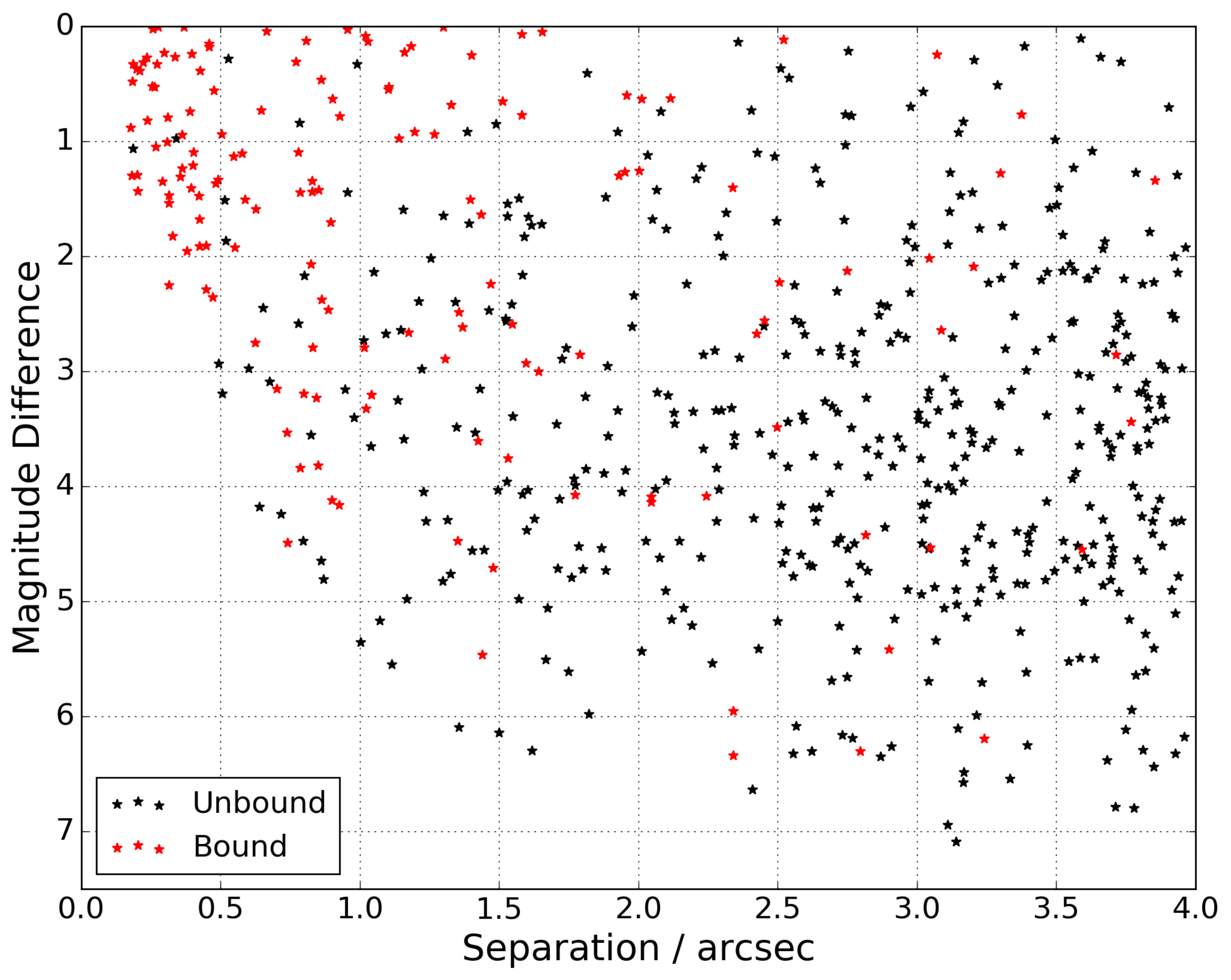}
\caption{Simulated Robo-AO survey using Galactic stellar models, described in Section$~\ref{sec:trilegal}$.  Nearby stars that are bound are plotted in red, and unbound asterisms are plotted in black.  Bound stars are likely to be found at small separations and near equal brightness to the target star.}
\label{fig:simulatedfield}
\end{figure}

\begin{figure}
\includegraphics[width=0.4
\paperwidth]{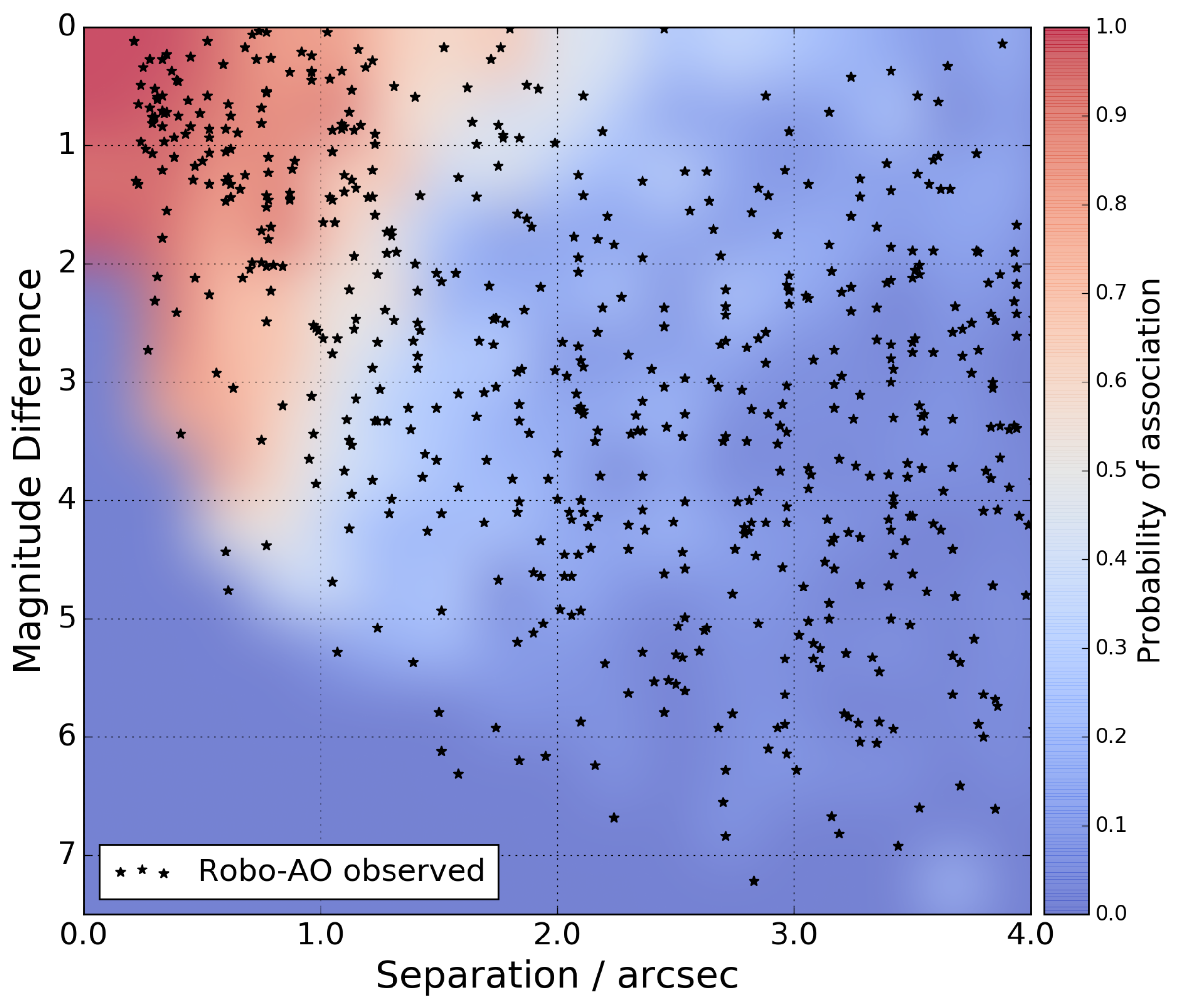}
\caption{Probability of association density map derived from simulated star fields from Galactic stellar models, as described in Section$~\ref{sec:trilegal}$.  Observed nearby stars to KOIs from the Robo-AO survey are overplotted. Most nearby stars detected within 1\arcsec of the planetary host star are likely bound. Stars outside of 2\arcsec are likely unbound, however still contaminate the \textit{Kepler} photometry resulting in incorrect planetary radii estimates.}
\label{fig:heatmap}
\end{figure}

We simulate the detectable systems with Robo-AO using our average image performance (see Section 3.5 of Paper IV) as a function of source brightness and a random variation caused by seeing. The properties (separation and contrast ratios) of the distribution of simulated nearby stars closely matches that of the observed nearby stars from Robo-AO observations. A simulated \textit{Kepler} field, with the number of nearby stars plotted equivalent to the number detected in the full Robo-AO KOI survey, is displayed in Figure$~\ref{fig:simulatedfield}$.  Using all ten simulated fields, we determine the probability of association for a given separation and contrast.  This probability density map is displayed in Figure$~\ref{fig:heatmap}$, with observed nearby stars to KOIs from the Robo-AO survey overplotted.

The results of these simulations generally agree with the previous simulations by \citet{horch14} and evidence from observations (displayed in Figure$~\ref{fig:observed_systems}$): most stars within 1\arcsec are expected to be bound, while wider separated companions with higher contrasts are likely unbound.

\subsubsection{Expected Giant Star Contamination}
\label{sec:giantcontamination}

It is conceivable that an unbound background giant star, observed near a KOI, has a distance estimate similar to the KOI resulting in a high probability of being bound.  This is a result of our assumption that the background stars are dwarf stars.  We estimate the number of expected giant stars being characterized as bound dwarf stars to the KOIs using the simulated \textit{Kepler} fields, discussed in Section$~\ref{sec:trilegal}$.  We perform our stellar characterization analysis on the ten simulated fields, described in Section$~\ref{sec:probassociation}$.  We find a probability of approximately 20\% that a single background giant star in the entire Robo-AO KOI survey will, if we assume it is a dwarf star, have an estimated probability of association with a planet host greater than 2$\sigma$ in our analysis. We therefore expect the impact of background giant star contamination on results in this work to be negligible.

\subsection{Galactic Latitude and Stellar Density}
\label{sec:stellardensity}

\begin{figure}
\centering
\includegraphics[width=245pt]{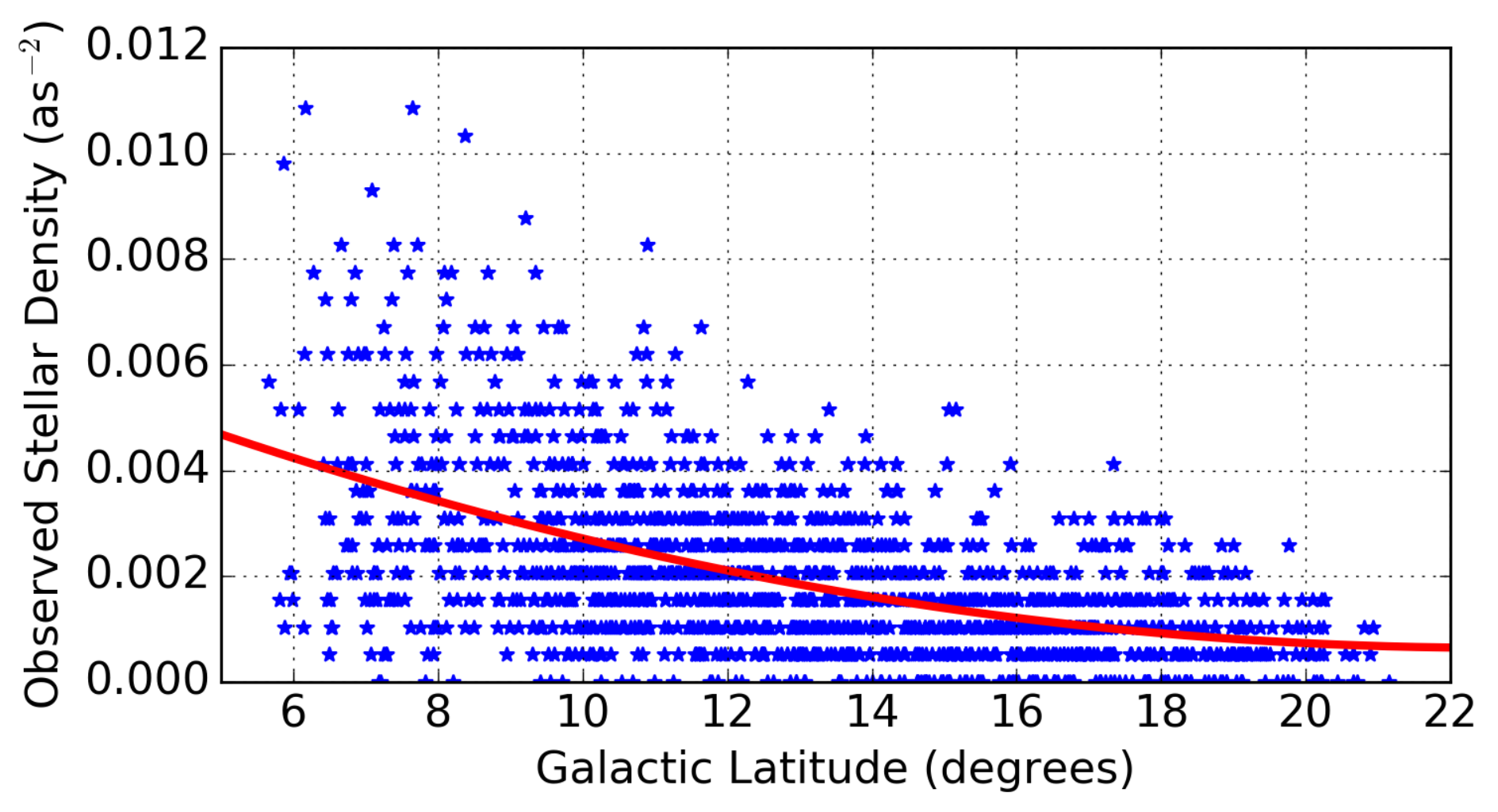}
\caption{Observed stellar densities in Robo-AO full-frame images within the \textit{Kepler} field as a function of Galactic latitude with quadratic fit.  Target stars and stars within 4\arcsec of the target star have been excluded.  The distribution reveals a negative correlation between stellar density and Galactic latitude.}
\label{fig:latitude}
\end{figure}

The location of the KOI within the \textit{Kepler} field may also impact the likelihood that an unbound star will be observed nearby. The large set of full frame Robo-AO images of KOIs, with a field of view 44\arcsec square, allow us to measure the observed stellar density over a statistically significant section of the sky within the \textit{Kepler} field of view. We counted stars within 2598 full-frame images, not including the target star or any stars within 4\arcsec of the target star, to determine the observed stellar density with Robo-AO as a function of Galactic coordinates. The typical depth of Robo-AO images, based on the image performance metrics described in \citet{ziegler18}, is 4-7 mag fainter than the KOI, equivalent to a V$\approx$20 star. The simulations described in Section$~\ref{sec:trilegal}$ suggest that the vast majority of stars outside 4\arcsec are unbound to the target star.  The observed stellar densities from the full frame Robo-AO images of KOI targets as a function of Galactic latitude are shown in Figure$~\ref{fig:latitude}$, with quadratic fitting line.

We find Pearson correlation coefficients (where a value of 0 signifies no linear correlation and a value or either 1 or -1 signifies total positive or negative linear correlation) of Galactic latitude and longitude to observed stellar densities of -0.53 and -0.03, respectively.  This suggests that, as expected, Galactic latitude is the primary variable in estimating local stellar density.  Indeed, the median Galactic latitude for KOIs with nearby stars is \textit{b}=11.2, approximately a degree and a half closer to the Galactic disk compared the median latitude of all KOIs (\textit{b}$_{med}$=12.7), while the difference in median Galactic longitude for the two populations is negligible.  Apart from a higher number of nearby unbound stars, KOIs with nearby stars may on average be found at lower Galactic latitudes due in part to higher intrinsic binarity rates of thick disk stars compared to thin disk stars \citep{chiba00, grether07}.

\begin{figure}
\centering
\includegraphics[width=245pt]{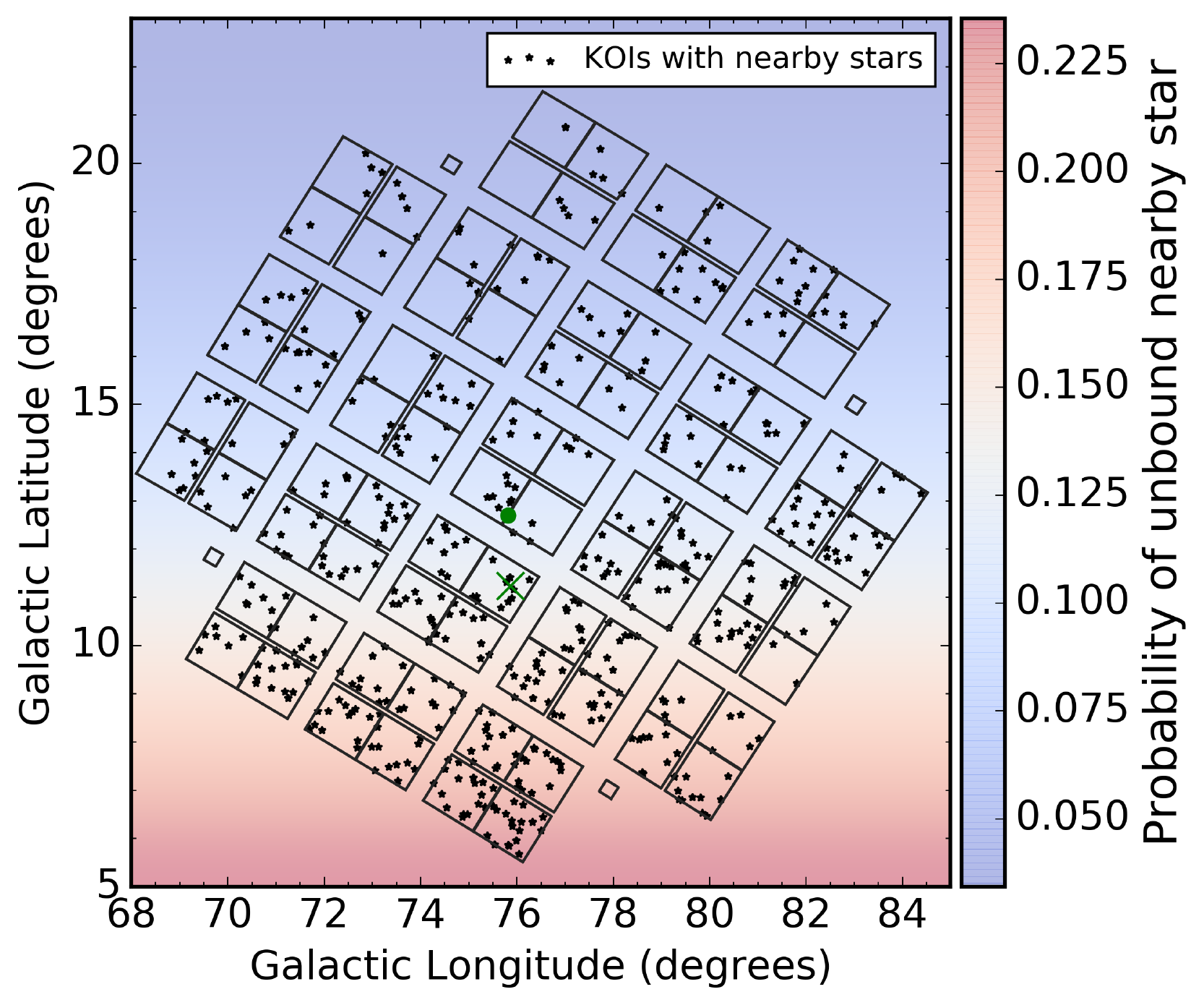}
\caption{Location on sky of KOIs with nearby stars from Paper I, II, III, and this work. A projection of the \textit{Kepler} field of view is provided for reference. The probability of an unbound star being found within 4\arcsec of the KOI is plotted, as determined from observed stellar densities with Robo-AO.  The median sky position of all observed KOIs and KOIs with nearby stars are plotted with a green circle and $\times$, respectively.  KOIs with nearby stars are on average closer to the Galactic disk.}
\label{fig:keplerfov}
\end{figure}

We use these stellar densities to then estimate the probability that an unbound star will, by chance, be within 4\arcsec of a KOI.  We plot the KOIs with nearby stars observed in the Robo-AO survey in Figure$~\ref{fig:keplerfov}$.  We also plot the probability that an unbound star will be observed nearby (within 4\arcsec) a KOI estimated from the quadratic fit to the observed stellar densities as a function of Galactic latitude. For the entire set of 3857 targets from the Robo-AO KOI survey, we would expect on average approximately 318 unbound stars to be observed within the same \textit{Kepler} pixel (separations within 4\arcsec) of the planetary hosts.

\section{Discoveries and Discussion}
\label{sec:Discussion}

In this section, we delve further into the full set of observations from the Robo-AO KOI survey to further explore the implications of stellar multiplicity on the planetary candidates (Section \ref{sec:implications}), and search for insight into the role that multiple stellar bodies play on planetary formation and evolution (Section \ref{sec:multiplicitystudy})

\subsection{Implications for \textit{Kepler} Planet Candidates}
\label{sec:implications}

When a close companion is detected near a KOI host star, there are several potential implications as, in general, it is not known which star is the source of the transit signal.  If the planet does indeed transit the purported target star, the consequences may be relatively mild: the planet's radius will be slightly larger than had previously been thought---at most by a factor of $\sqrt{2}$ in the case of an equal-brightness companion \citep{ciardi15}.  If the eclipsed star is a faint companion, however, the radius of the eclipsing object may be many times larger, potentially turning a small planet into a giant planet or a planet into a false-positive eclipsing binary star. In Paper IV, we found that for a scenario in which each planet is equally likely to be hosted by the primary or secondary star, the planetary radii will increase on average by a factor of 2.18. The large number of faint unassociated stars likely inflates this factor; for just systems with stars within 1\arcsec, the planetary radii will increase by a factor of 1.54.

The radii estimates of the primary stars in Paper IV were from \textit{Kepler} stellar catalog \citep{dr25}. As the properties of most of the host stars in the \textit{Kepler} stellar catalog are based on broad-band photometry assuming that they are single, the derived stellar radius of the primary star may well be incorrect if the system actually contains multiple stars. We estimated the radii of the secondary stars in Paper IV under the assumption that they were bound to the primary. We then used the stellar radius of an appropriately faint star (based on the visible-band contrast with respect to the primary) in the Dartmouth stellar models \citep{dotter08}. 

In this paper, we use multi-band photometry from the resolved systems to estimate the spectral type of 145 KOIs and nearby stars, as described in Section$~\ref{sec:probassociation}$. We use the spectral types to estimate the stellar radius of each star, and correct the estimated radius of planets around those KOIs using the methodology detailed in Section$~\ref{sec:updated_radii}$. The corrected planetary radii are available in Table$~\ref{tab:radii}$. 

We find radius correction factors for these 145 KOIs of 1.06, 4.65, and 2.86 for the scenarios in which all planets orbit the primary, all planets orbit the secondary, and planets are equally likely to orbit the primary and secondary, respectively. As discussed in Paper IV, the latter two scenarios are not likely due to the large number of unbound stars in our sample. Assuming the planets are equally likely to orbit these unbound stars results in a large number of gas giant planets, which are inherently rare in the galaxy \citep{howard12}. If we use just the likely bound stars (as determined in Section$~\ref{sec:probassociation}$), the planetary radii will increase by factors of 1.08, 2.47, and 1.77 for the scenarios in which all planets orbit the primary, all planets orbit the secondary, and planets are equally likely to orbit the primary and secondary, respectively.

It is believed that the transition from a rocky planet to a planet with a large gaseous envelope begins relatively sharply at 1.6 $R_{\oplus}$ \citep{rogers15}. In our set of 145 KOIs, we find 10 planet candidates, each initially believed to be rocky (R$_{p,0}$$<$1.6$ R_{\oplus}$), are likely not rocky, with corrected radii larger than 1.6 $R_{\oplus}$, whether orbiting either the primary or secondary star. All but two of these (KOI-2380.01 and KOI-4713.01) were previously determined to not be rocky regardless of host star in Paper IV. We also find that most (72 of 81) of the planetary candidates with initial radius estimates less than 1.6$R_{\oplus}$, consistent with rocky composition, would likely not be rocky if hosted by the secondary star.

\subsubsection{Impact on the Planet Radius Gap}

The California \textit{Kepler} Survey (CKS) recently released updated stellar parameters \citep{cks} several times more precise than those derived from photometry in the \textit{Kepler} input catalog. \citet{fulton17} filtered the 2025 KOIs in the CKS sample to remove false positives, giant stars, low-impact parameter planets, faint stars, long-period planets, and non-solar type stars, resulting in a set of 900 KOIs. Within this filtered set of KOIs, they detected a gap in the planetary radius distribution for planets with radii between $1.5-2.0 R_{\oplus}$. This gap has been interpreted as an ``evaporation-valley'' between a population of rocky super-Earths and a population of sub-Neptunes with thick atmospheric envelopes \citep{owen17}

The depth of the radius gap can be quantified using the metric $V_{A}$, derived in \citet{fulton17}, which is the ratio of the number of planets in the bottom of the gap (1.64-1.97 $R_{\oplus}$) to the average number in the peaks immediately outside of the gap (1.22-1.44 $R_{\oplus}$ and 2.16-2.62 $R_{\oplus}$). Smaller values of $V_{A}$ denote a deeper gap. The filtered CKS sample without radius corrections has a gap depth of $V_{A}$=0.483. 

\begin{figure}
\centering
\includegraphics[width=245pt]{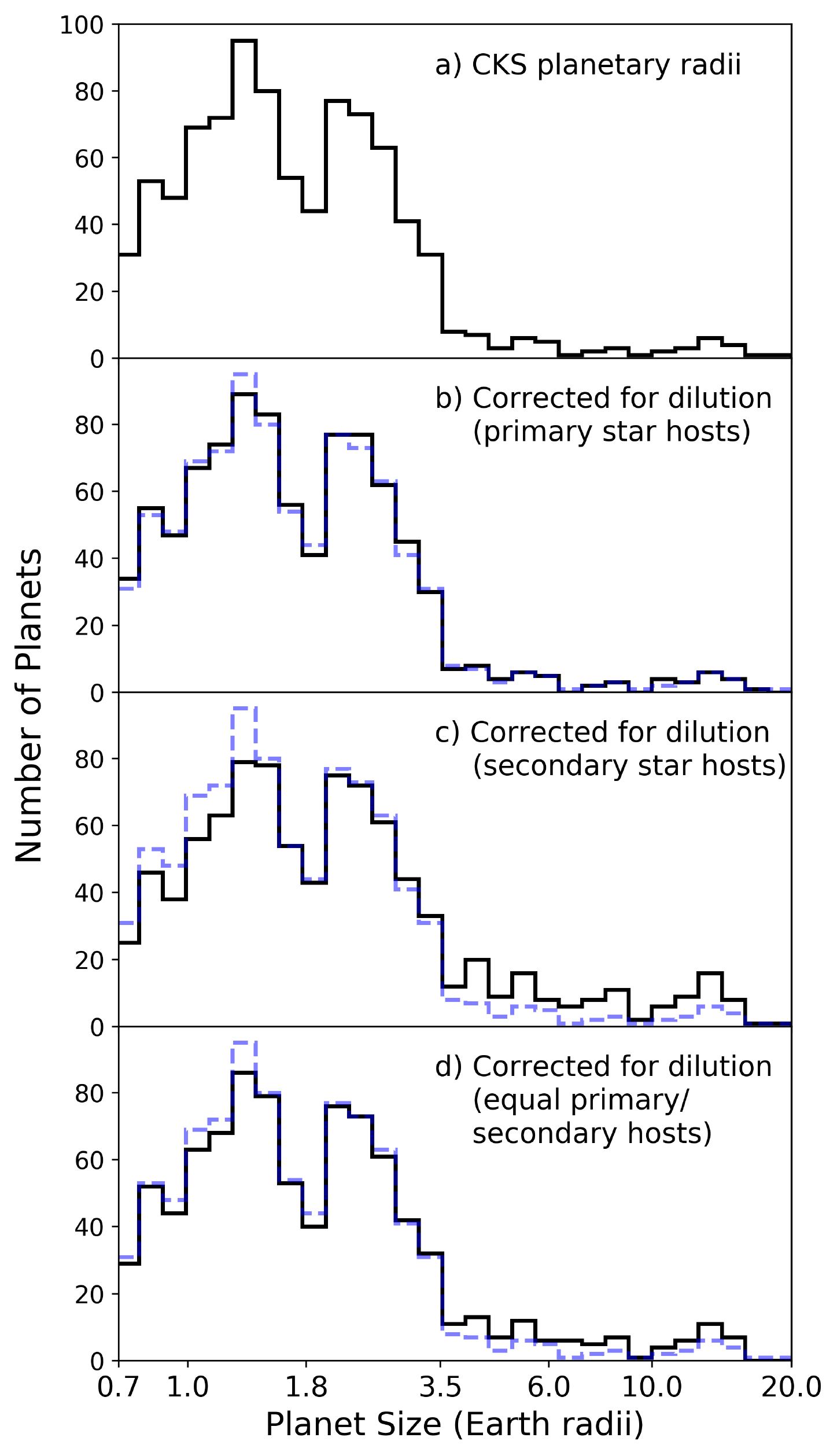}
\caption{The radius distribution of planets using stellar parameters from the California \textit{Kepler} Survey \citep{cks}. In Panel \textbf{(a)}, the planetary radius distribution after successive filters have been applied, revealing a gap in the radius distribution for small planets \citep{fulton17}. Panels \textbf{(b)-(d)} show the distribution after radius corrections to account for dilution from nearby stars detected by Robo-AO, assuming that all planets orbit the primary star, all planets orbit the secondary star, and planets are equally likely to orbit the primary or secondary star, respectively. The original CKS distribution from Panel \textbf{(a)} is overplotted with a dashed blue line for comparison.}
\label{fig:fultongap}
\end{figure}

In the Robo-AO survey, we detected nearby stars to 168 of the 900 KOIs (18.7$\%$) used in their analysis. We correct the radius of planets due to dilution from contaminating stars, using the corrected radii listed in Table$~\ref{tab:radii}$ for the 145 systems characterized in Section$~\ref{sec:probassociation}$. For systems without multi-band photometry, the radius corrections from Paper IV were used. The resulting distributions are shown in Figure$~\ref{fig:fultongap}$. We find that, if all planets in systems with detected nearby stars orbit the primary star, the gap deepens slightly, with $V_{A}$=0.463. If instead, all planets in multiple systems orbit the secondary star, the gap depth is reduced to $V_{A}$=0.521. In perhaps the more likely scenario, where planets are equally likely to orbit the primary or secondary stars, the gap is slightly deeper than with the original radius estimates, at $V_{A}$=0.468.

\begin{figure}
\centering
\includegraphics[width=245pt]{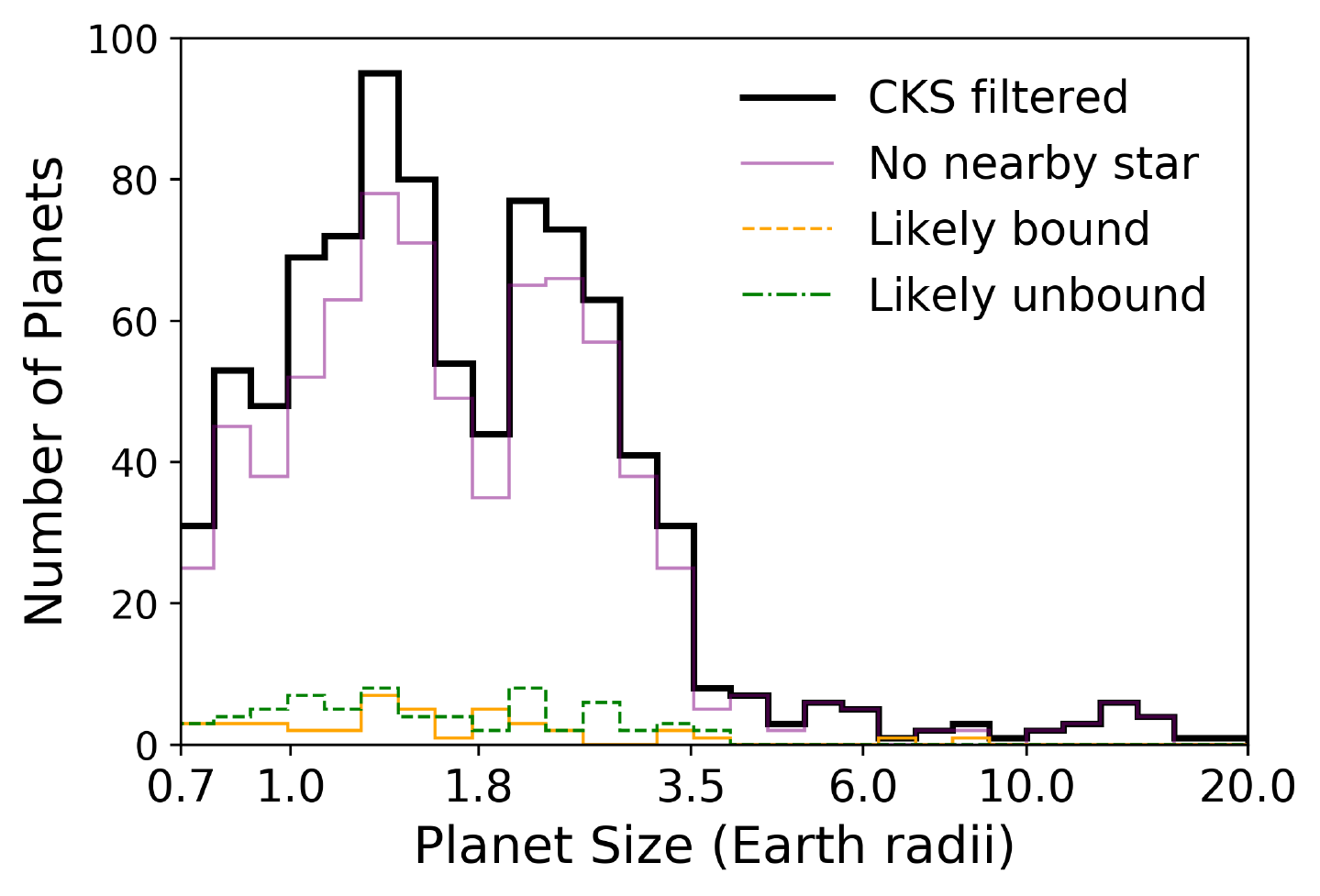}
\caption{The full distribution of planetary radii \citep{fulton17} from CKS is plotted in black, with the distribution from systems without detected nearby stars in purple. The radius distribution of the planets in systems with detected nearby stars that are likely bound and unbound are plotted in orange and green. Neither distribution has been corrected due to dilution from nearby stars. }
\label{fig:fultongap2}
\end{figure}

We can also compare the planetary radius distribution in systems with likely bound or unbound stars to the full CKS radius distribution, shown in Figure$~\ref{fig:fultongap2}$. We find a significant fraction of the large planets in the CKS sample have detected nearby stars (see Section$~\ref{sec:hotjupiters}$). We also find that the radius distributions of the set of all small planets (R$_{p}<3.5 R_{\oplus}$) and small planets in systems with nearby stars are statistically similar (p-value$>$0.85).

There is, however, a possible disparity in the radius distribution of planets in single and multiple star systems. For small planets in likely bound systems, no gap in the radius distribution is apparent. A Kolmogrov-Smirnov test reveals that the radius distributions (without dilution corrections) of the 781 small planets from the CKS sample in systems without nearby stars and small planets in systems with likely bound stars (44 planets, as determined in Section$~\ref{sec:probassociation}$) are significantly dissimilar with 95\% confidence (p-value of 0.04).\footnote{The two distributions are also significantly dissimilar (p-value of 0.05) when including dilution corrections under the assumption that either the primary or secondary star is equally likely to be the planet host.} This may be a result of small-number statistics as there are few known binaries hosting small planets. This disparity could, however, be evidence of the impact companion stars have on the bimodal distribution of small planets. Robo-AO continues to characterize multiple KOI systems to determine boundness, and we will revisit this discussion in future papers.

\subsection{Stellar Multiplicity and \textit{Kepler} Planet Candidates}
\label{sec:multiplicitystudy}
Combining the entire Robo-AO KOI survey, we detect 620 nearby stars around 569 planetary candidate hosts from 3857 targets, for an overall multiplicity fraction of 14.7\%$\pm$0.6\% within the detectability range of our survey ($\sim$0$\farcs$15--4$\farcs$0, $\Delta$m$\le$6).  With this large dataset we continue our search that began in Paper I and was updated in Paper III for broad-scale correlations between the observed stellar multiplicity and planetary candidate properties. Such correlations provide an avenue to constrain and test planet formation and evolution models. 

In addition to using a data set nearly twice as large as in our previous analysis, our stellar characterization of the target and nearby stars and analysis of physical association probabilities, described in Section$~\ref{sec:probassociation}$, allow us to attempt to remove the diluting impact of unbound nearby stars and strengthen any true correlation discovered.  In addition, we use the improved stellar parameters from CKS to search for correlations in multiplicity and stellar properties such as metallicity and temperature, that were formerly not well constrained.

Unless noted, all stellar and planetary properties for the KOIs in this section were obtained from the cumulative planet candidate list at the NASA Exoplanet Archive\footnote{\url{http://exoplanetarchive.ipac.caltech.edu/}} and have not been corrected for possible dilution due to the presence of nearby stars outside the \textit{Kepler} input catalog. The planet catalog does not correct for dilution from stars detected from high-resolution imaging as these observations are typically only available for KOIs and could bias occurrence rate estimates \citep{thompson18}.

\subsubsection{Stellar Multiplicity and Multiple-planet Systems Revisited}

It is thought that the impact of a stellar companion to the planetary host star should perturb multiple planet systems, leading to fewer observed multiple transiting systems. \citet{wang14} and \citet{picogna15} suggests that perturbations from the companion star will change the mutual inclination of planets in the same system. Planets in nearby orbits are also expected to perturb each other \citep{rasio96, wang15a}, possibly leading to planets being ejected out of the system \citep{xie14}.

\begin{figure}
\includegraphics[width=0.4
\paperwidth]{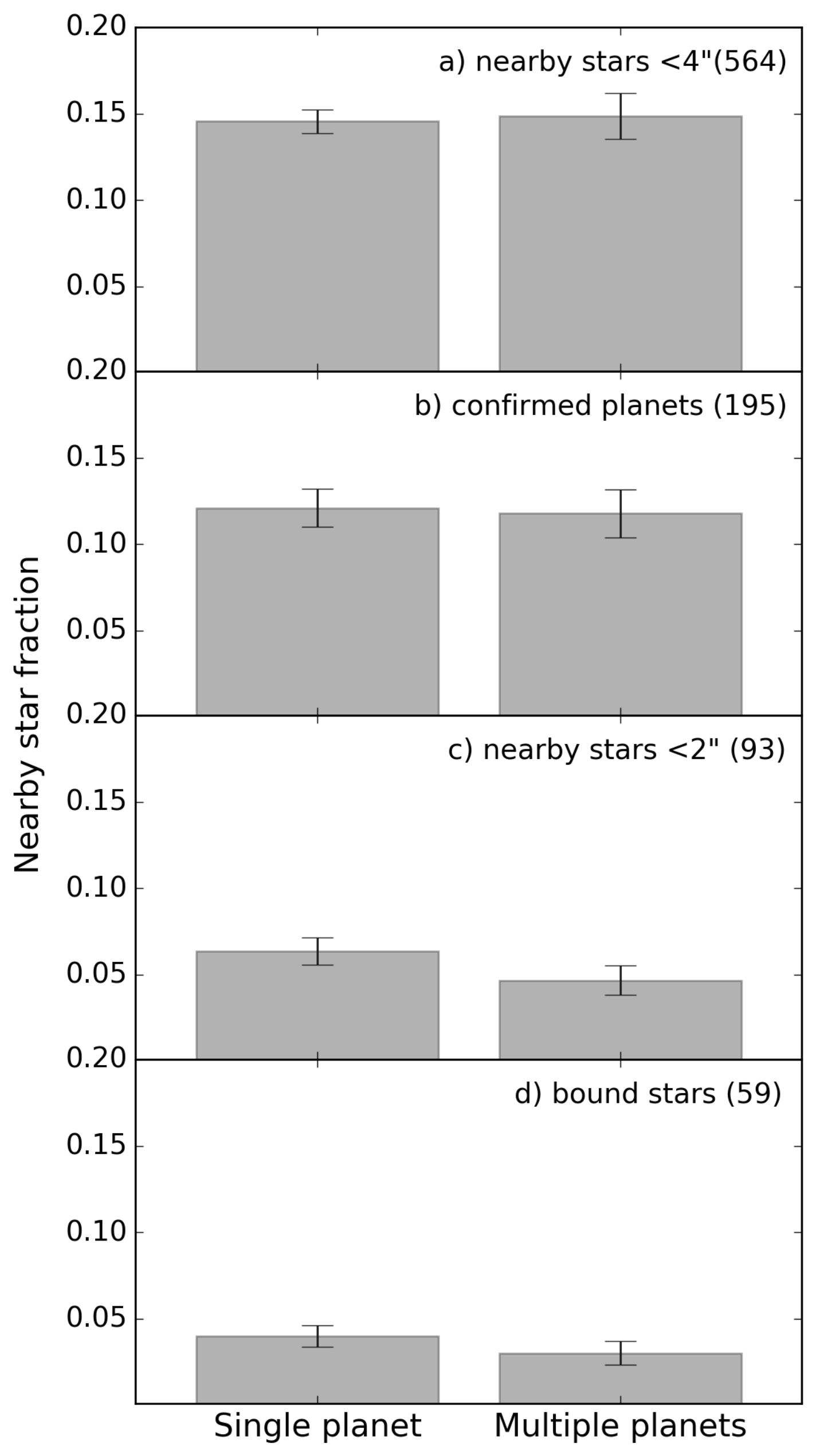}
\caption{\textbf{(a)} Nearby star fraction within 4\arcsec of KOIs hosting single- and multiple-planetary systems. Panels \textbf{(b)-(d)} show these fractions after successive cuts to \textbf{(b)}: remove systems with unconfirmed planets; \textbf{(c)}: remove systems with nearby stars at separations greater than 2\arcsec; \textbf{(d)}: remove systems with stars shown to be likely unbound from observations (see Section$~\ref{sec:probassociation}$), and weight systems with nearby stars whose association has not been studied by the probability of association based on the separation of the nearby star from the primary star derived from observations (see Figure$~\ref{fig:observed_systems}$).  The number of systems with nearby stars remaining after each successive cut is annotated in the upper right corner of each panel. In panel \textbf{(d)}, the weights of the systems with nearby stars without association determination were summed with the number of observed likely bound systems and rounded to the nearest whole number.}
\label{fig:numplanets}
\end{figure}

We searched for evidence of a disparity in stellar multiplicity between the two planetary populations in Paper I and Paper III. In Paper I, we found single-planet systems exhibiting a slightly higher nearby star fraction. With several times more targets used in the analysis in Paper III, we found a slightly higher nearby star fraction for the multiple-planet systems. Combining all KOI targets, we again find little difference between the two populations (displayed in panel \textit{a} in Figure$~\ref{fig:numplanets}$): a Fischer exact test gives 87\% probability the two populations are drawn from the same distribution.  

The full set of KOI targets is likely highly diluted, however, by false positive planets \citep{morton11, fressin13} and unbound nearby stars (see Section $~\ref{sec:probassociation}$). In determining whether a nearby star is bound or not, we first refer to the results of the photometric-distance estimates. If the result was of this analysis was uncertain, or if the system had not been observed yet, we weight the system using the probability of association as determined in Section $~\ref{sec:trilegal}$, based on the separation and contrast of the star with respect to the primary star.  We therefore perform cuts to the set of KOI targets, removing candidate planets, systems with nearby stars at greater than 2\arcsec separation, and likely unbound stars, in an attempt to reduce these effects, shown in Figure$~\ref{fig:numplanets}$.  We find after all successive cuts, with confirmed planets with likely bound nearby stars, single-planet systems have slightly higher nearby star fraction rate than multiple-planet systems: 4.0\%$\pm$0.6\% and 3.0\%$\pm$0.7\%, respectively. A Fischer exact test gives two-thirds probability (66.5\%) that the two populations are indeed disparate.

It is not clear if this low-significance result is evidence of the disturbing impact of stellar companions on planetary systems. Other factors may result in a higher than expect binarity fraction of multiple planet systems, however. Companion stars can cause orbital migration with Kozai oscillations \citep{fabrycky07}, shifting multiple planets in the same system to shorter periods where \textit{Kepler} has higher sensitivity to transit events.  Binary stars that form together also may have mutually inclined protoplanetary disks \citep{muller12}, leading to separate transiting planetary systems around each star.

\subsubsection{Stellar Multiplicity and Close-in Planets Revisited}
\label{sec:hotjupiters}
It is hypothesized that the presence of a stellar companion will greatly influence the properties and architecture of planetary systems. Observational evidence suggests that planetary formation is suppressed in close binaries, resulting in a fifth of all solar type stars being unable to host planets because of stellar interactions \citep{kraus16}. Perturbations from the nearby star are thought to drive planets that form at large separations inward to low-period orbits \citep{fabrycky07, katz11, naox12}, with smaller planets more susceptible due to weak planet-planet dynamical coupling \citep{wang15a}. Interactions between planets within the same system, often caused by orbital migration caused by stellar companions, are thought to eject small planets at a greater rate than giant planets \citep{xie14}. We would expect then a correlation between binarity and planetary period for different sized planets.

\begin{figure}
\includegraphics[width=0.4
\paperwidth]{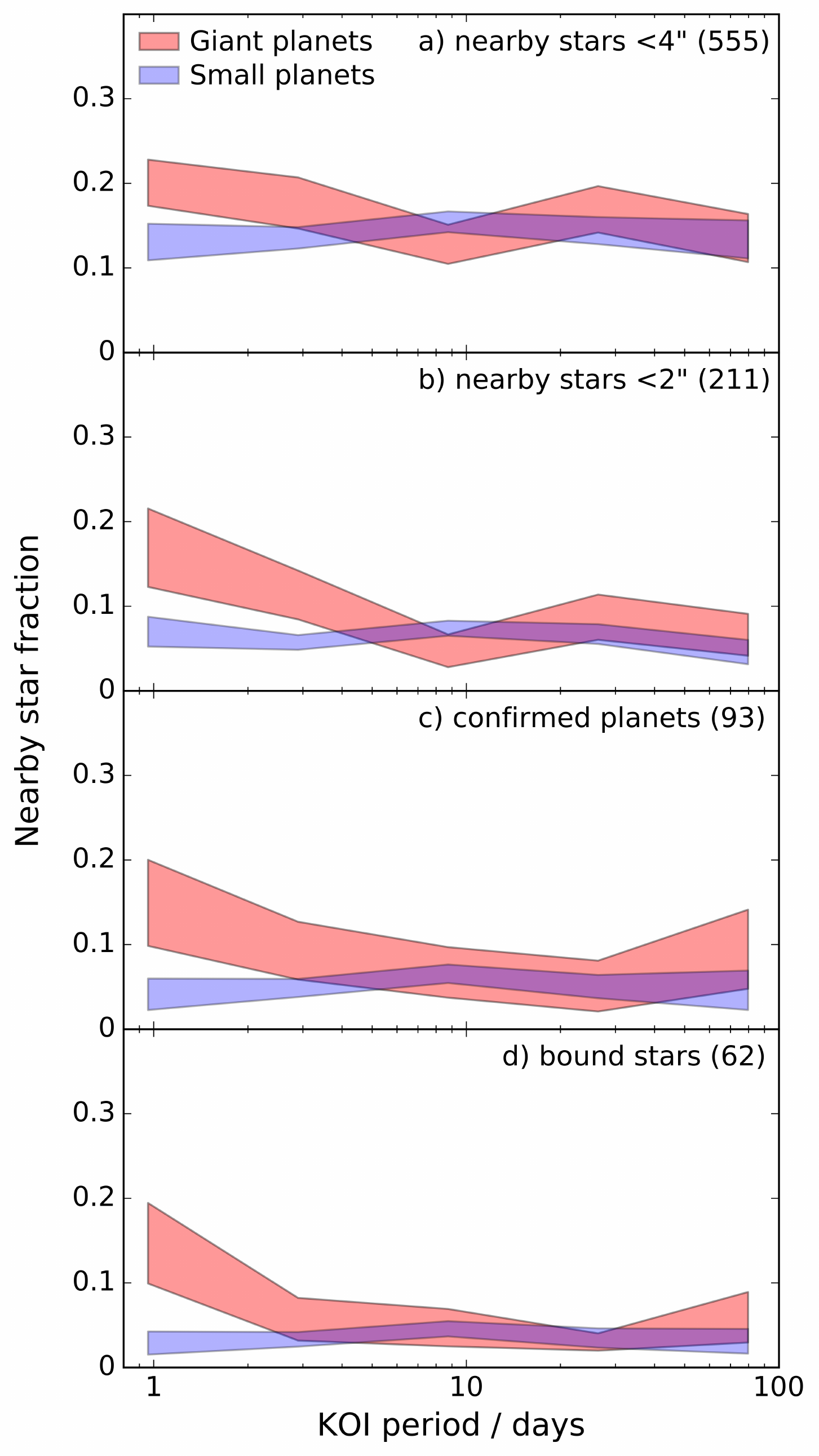}
\caption{\textbf{(a)} 1$\sigma$ uncertainty regions for the nearby star fraction as a function of KOI period for two different planetary populations. Panels \textbf{(b)-(d)} show these regions after successive cuts to \textbf{(b)}: remove systems with nearby stars at separations greater than 2\arcsec; \textbf{(c)}: remove systems with unconfirmed planets; \textbf{(d)}: remove systems with stars shown to be likely unbound from observations (see Section$~\ref{sec:probassociation}$), and weight the systems with nearby stars not characterized with multi-band photometry based on their separation and magnitude difference with respect to the primary star (see Figure$~\ref{fig:observed_systems}$).  The number of systems with nearby stars remaining after each successive cut is annotated in the upper right corner of each panel. In panel \textbf{(d)}, the weights of systems with uncharacterized nearby stars were summed with number of observed likely bound systems and rounded to the nearest whole number.}
\label{fig:hotjupiters}
\end{figure}

In our analysis in Paper III, we found little evidence for a disparity in nearby star fraction for giant (R $>$ 3.9 R$_\oplus$) and small (R $<$ 3.9 R$_\oplus$, Neptune radius) planets at short or long periods.  With the combined data set of the Robo-AO KOI survey, we find a low-significance increase in nearby star fraction for giant planets at short periods compared to small planets (see panel \textbf{a} in Figure$~\ref{fig:hotjupiters}$).

We expect that our sample is heavily diluted, however, by false positive planets (hot Jupiters are expected to have a higher than average false positive rate \citealt{santerne12}) and unassociated nearby stars, as discussed in Section $~\ref{sec:probassociation}$.  We therefore remove contaminating systems in search of any possible underlying correlation.  These cuts are as follows (resulting distributions are shown in Figure$~\ref{fig:hotjupiters}$): remove nearby stars at separations greater than 2\arcsec, a region where most stars are highly likely to be unbound; remove unconfirmed planets; remove systems that have been shown to be likely unbound from observations (see Section $~\ref{sec:probassociation}$), and weight systems whose probability of association has not been studied by the percent likelihood of being bound based on their separation from the host star, as determined from observations.

After successive cuts, we find that giant and small planets on 1-3 day orbits have a binarity rate of 12.8\%$^{+5.6\%}_{-2.8\%}$ and 2.4\%$^{+1.8\%}_{-0.9\%}$\footnote{Errors for both populations are based on Poissonian statistics \citep{burgasser03}}, respectively, a 2.6$\sigma$ discrepancy.  No other period range shows a significant difference in binarity rate between the two populations.

This result agrees with the NIR survey of \citet{ngo15}, that found hot Jupiter hosts are twice as likely as field stars to be found in multiple star systems, with a significance of 2.8$\sigma$.  They, however, find that 51\% of hot Jupiters are hosted by stars with stellar companions; the discrepancy in the binarity fractions found in the two surveys likely is a result of differing observational methods and limits.  The binarity fraction for hot Jupiters in this work does agree with that found by \citet{roell12} of 12\% using binary catalogues based primarily on seeing-limited observations.

\subsubsection{Stellar Multiplicity Rates and Host-star Temperature Revisited}

\begin{figure}
\centering
\includegraphics[width=245pt]{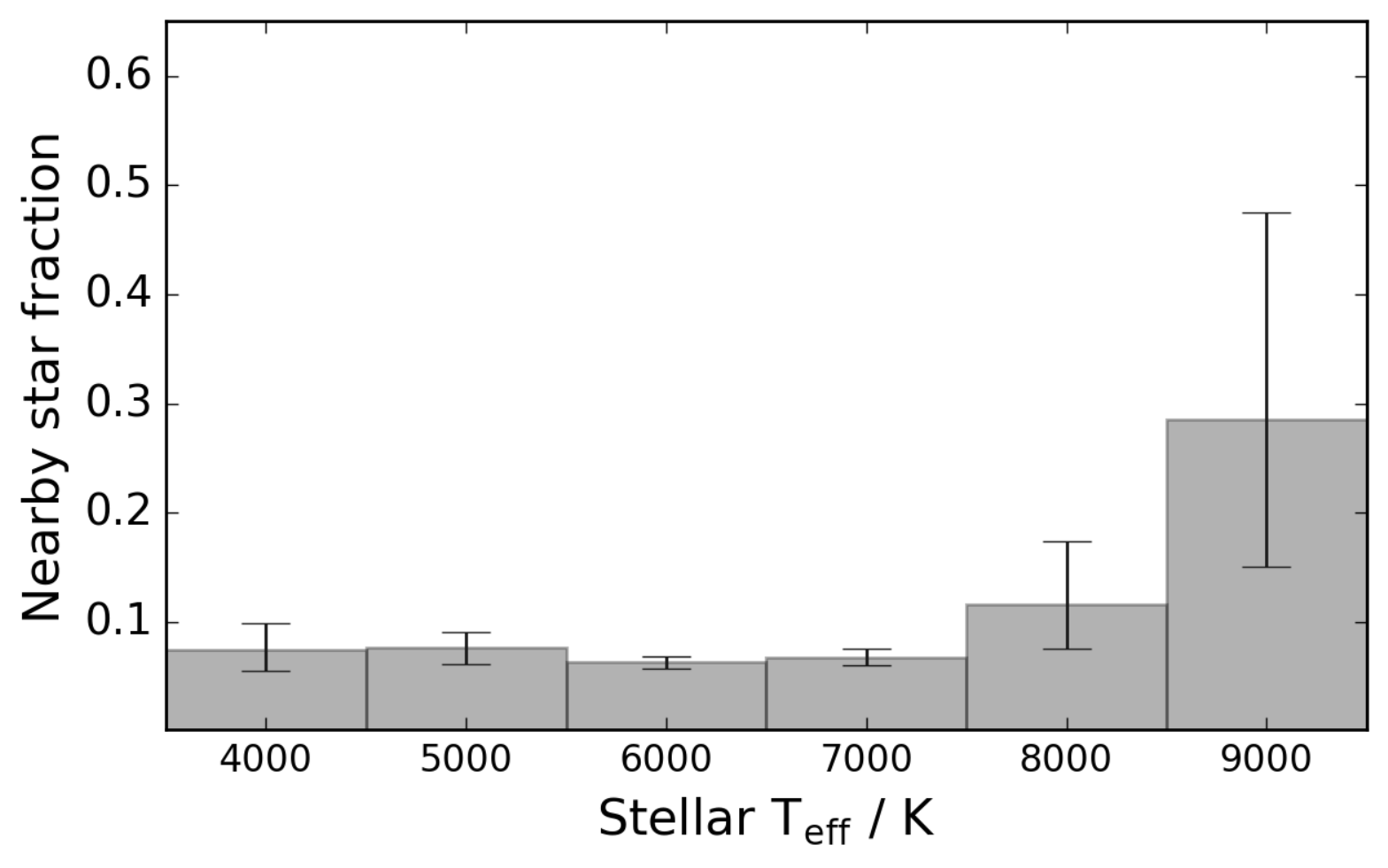}
\caption{Fraction of KOIs with detected nearby ($\le$2\arcsec) stars as a function of stellar effective temperature.}
\label{fig:teffbinaryfrac}
\end{figure}

\begin{figure}
\centering
\includegraphics[width=245pt]{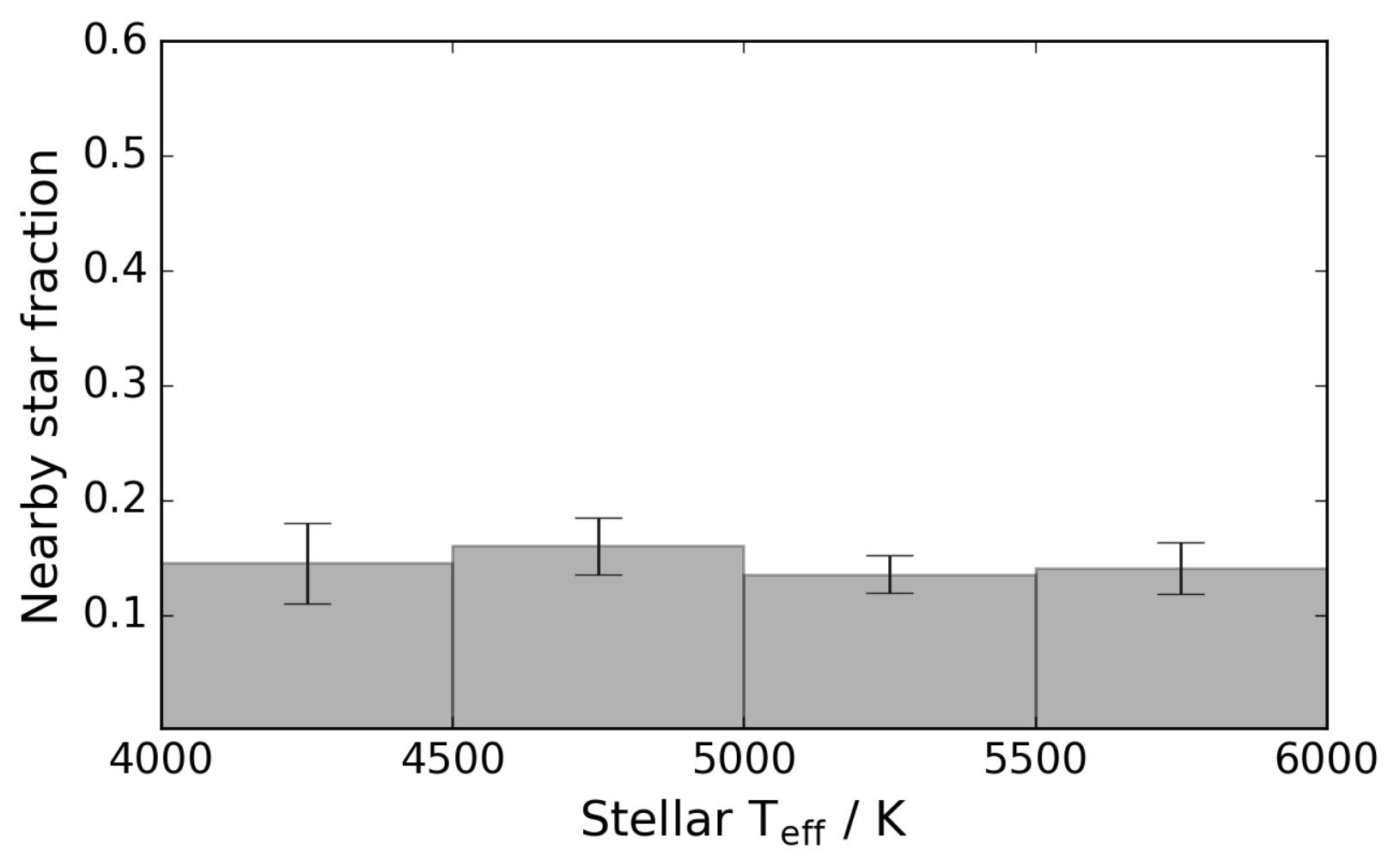}
\caption{Fraction of KOIs with detected nearby ($\le$4\arcsec) stars as a function of stellar effective temperature, using estimates from the California \textit{Kepler} Survey \citep{cks}.}
\label{fig:ckstemps}
\end{figure}

We found that KOIs follow the correlation between multiplicity and stellar mass and temperature observed in field stars \citep{duchene13} in Paper III. Many of these field stars likely host their own non-aligned planets. Restricting our sample to the likely bound nearby stars with separations less than 2\farcs0 (as discussed in Section$~\ref{sec:probassociation}$), we find that the trend remains for the entire set of observations from the Robo-AO KOI survey, as seen in Figure$~\ref{fig:teffbinaryfrac}$. The majority of stars in the CKS survey are solar-type, with effective stellar temperatures between 4500 and 6500 K. The trend relating multiplicity and effective temperature is expected to be negligible in that compact range of stellar temperatures, and indeed, no significant trend is apparent as seen in Figure$~\ref{fig:ckstemps}$.

\subsubsection{Stellar Multiplicity and Metallicity of KOIs}

The relationship between stellar multiplicity and metallicity is not well understood.  Early studies suggested that metal-poor stars possessed fewer stellar companions \citep{kopal59, jaschek59, batten73, abt87}. A previous Robo-AO survey \citep{ziegler15} found low-metallicity cool subdwarf stars had binary fractions $\sim$3$\times$ lower than similar solar-metallicity dwarf stars. More recent studies, however, have suggested that multiplicity rates decrease with metallicity \citep{carney87, carney94}.  In particular, \citet{grether07} found a $\sim$2$\sigma$ anti-correlation between metallicity and companion stars.  Planetary systems do seem to occur more frequently in metal-rich stars \citep{fischer04, grether07}.

We can use the multiplicity fraction of planet candidate hosting stars as a function of metallicity to determine how these stars compare to field stars.  For this analysis, we use the precise metallicity ([Fe/H]) estimates from the CKS \citep{cks} which have typical uncertainties of 0.05 dex.

We may expect to see a correlation between metallicity and nearby star fraction rate in the full set of KOIs since our sample likely has a high number of unbound stars (see Section$~\ref{sec:probassociation}$. A higher fraction of these nearby stars are from low Galactic latitudes where the observed stellar density is greater (see Section$~\ref{sec:stellardensity}$). \citet{grether07} found that stars at low \textit{b}, citizens of the Galactic thick disk which have higher average metallicity \citep{ishigaki12}, shows a $\sim$4 times higher binary fraction than halo stars.

\begin{figure}
\includegraphics[width=0.4
\paperwidth]{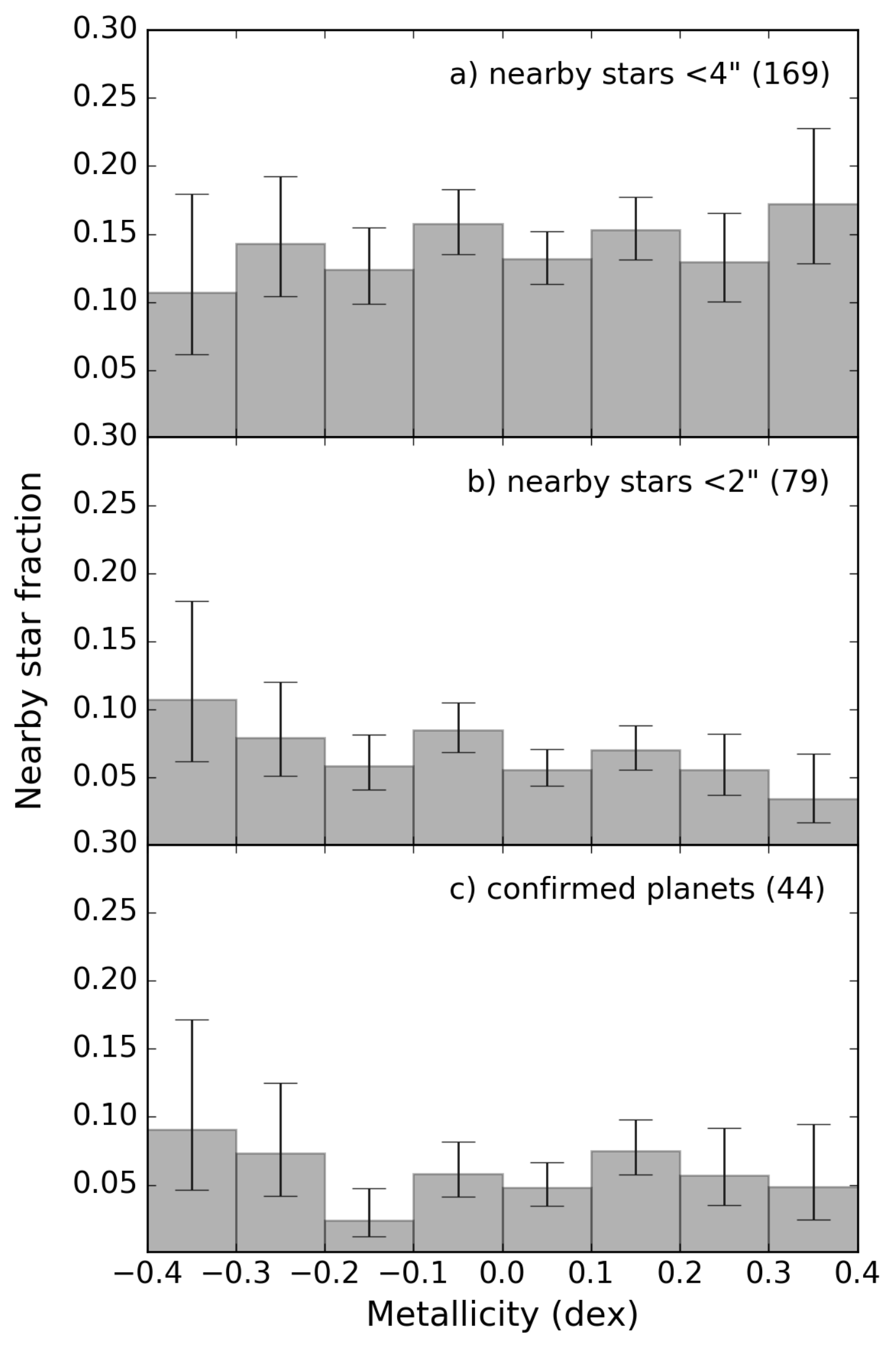}
\caption{\textbf{(a)} Nearby star fraction as a function of KOI metallicity ([Fe/H]) using CKS estimates \citep{cks}. Panels \textbf{(b)} and \textbf{(c)} show these regions after successive cuts to \textbf{(b)}: remove systems with nearby stars at separations greater than 2\arcsec; \textbf{(c)}: remove systems with unconfirmed planets. The number of systems with nearby stars remaining after each successive cut is annotated in the upper right corner of each panel.}
\label{fig:metallicity}
\end{figure}

For the set of all KOIs, the nearby star fraction within 4\arcsec visually appears to correlate slightly with metallicity, as shown in panel \textbf{a} in Figure$~\ref{fig:metallicity}$.  A Fisher exact test suggests however with high probability ($\sim$99\%) that sub- and super-solar metallicity KOIs are similar populations, with nearby star fraction rates of 14.1\%$\pm$1.7\% and 14.1\%$\pm$1.4\%, respectively.

If we remove any systems with nearby stars at separations greater than 2\arcsec, which are likely to be unbound as discussed in Section$~\ref{sec:probassociation}$, we find that the nearby star fraction rate slightly decreases as metallicity increases, shown in panel \textbf{b} of Figure$~\ref{fig:metallicity}$.  With the decreased separation limit, sub-solar and super-solar metallicity KOIs have nearby star fraction rates of 7.7\%$\pm$1.3\% and 5.9\%$\pm$0.9\%, respectively; a Fisher exact test gives a 28\% probability that the two populations are distinct.

Finally, when we limit the sample to confirmed planets, shown in panel \textbf{c} of Figure$~\ref{fig:metallicity}$, no significant trend is apparent between stellar binarity of planet hosting stars and stellar metallicity.  Now, a Fischer exact tests suggests with 87\% probability that sub- and super-solar are similar stellar populations, binarity rates of 5.4\%$\pm$1.3\% and 5.9\%$\pm$1.2\%, respectively.

\section{Conclusion}
\label{sec:conclusion}

Combining the data sets from the complete Robo-AO KOI survey, we found 630 nearby stars around 569 planetary candidate hosts, from a target list of 3857 KOIs, implying a nearby star fraction rate of 14.7\%$\pm$0.6\% within the Robo-AO detectability range (separations between $\sim$0$\farcs$15 and 4$\farcs$0 and $\Delta$m$\le$6).

We used galactic stellar models and the observed stellar density to estimate the number and properties of unbound stars detected in the survey. We characterized the primary and nearby star in 145 KOI systems reobserved with Robo-AO in multiple visible bands.  We quantified the probability of association for these systems, and derive corrected planetary radii for planetary candidates within these systems.  With the sample in this work and previously published datasets, we find that most stars within 1\arcsec of a KOI are likely bound.

We corrected the estimated planetary radii for planets within the 145 reobserved systems. We found that the planetary radii in likely bound systems will increase on average by a factor of 1.77, if either star is equally likely to host the planet. We found that the gap detected in the radius distribution of small planets is robust to the impact of dilution. We also found a low-significance disparity between the radius distribution of small planets in single and binary systems.

We found that giant planets at low periods are several times more likely be found in systems with stellar companions than other planets. We found that single and multiple planet systems are equally likely to orbit in binary star systems. We found that KOIs follow trends observed in field stars with respect to the relationship between stellar multiplicity and stellar effective temperature and metallicity.


\section*{Acknowledgements}
This research is supported by the NASA Exoplanets Research Program, grant $\#$NNX 15AC91G. C.Z. and W.H. acknowledge support from the North Carolina Space Grant consortium.  C.B. acknowledges support from the Alfred P. Sloan Foundation. T.M is supported by NASA grant $\#$NNX 14AE11G under the Kepler Participating Scientist Program.

We thank the observatory staff at Kitt Peak for their efforts to assist Robo-AO KP operations and are grateful to the Palomar Observatory staff for their support of Robo-AO on the 1.5-m telescope. 

The Robo-AO instrument was developed with support from the National Science Foundation under grants AST-0906060, AST-0960343, and AST-1207891, IUCAA, the Mt. Cuba Astronomical Foundation, and by a gift from Samuel Oschin.

The Robo-AO team thanks NSF and NOAO for making the Kitt Peak 2.1-m telescope available. Robo-AO KP is a partnership between the California Institute of Technology, the University of Hawai‘i, the University of North Carolina at Chapel Hill, the Inter-University Centre for Astronomy and Astrophysics (IUCAA) at Pune, India, and the National Central University, Taiwan. The Murty family feels very happy to have added a small value to this important project. Robo-AO KP is also supported by grants from the John Templeton Foundation and the Mt. Cuba Astronomical Foundation. 

Some data are based on observations at Kitt Peak National Observatory, National Optical Astronomy
Observatory (NOAO Prop. ID: 15B-3001), which is operated by the Association of Universities for Research in Astronomy (AURA) under cooperative agreement with the National Science Foundation.

Some of the data presented herein were obtained at the W.M. Keck Observatory, which is operated as a scientific partnership among the California Institute of Technology, the University of California and the National Aeronautics and Space Administration. The Observatory was made possible by the generous financial support of the W.M. Keck Foundation. We recognize and acknowledge the very significant cultural role and reverence that the summit of Maunakea has always had within the indigenous Hawaiian community. We are most fortunate to have the opportunity to conduct observations from this mountain.

This research has made use of the Exoplanet Follow-up Observation Program website, which is operated by the California Institute of Technology, under contract with the National Aeronautics and Space Administration under the Exoplanet Exploration Program

This research has made use of the NASA Exoplanet Archive, which is operated by the California Institute of Technology, under contract with the National Aeronautics and Space Administration under the Exoplanet Exploration Program.

{\it Facilities:} \facility{PO:1.5m (Robo-AO)}, \facility{KPNO:2.1m	(Robo-AO)}, \facility{Keck:II (NIRC2-LGS)}



\bibliography{references}

\appendix


\clearpage
\LongTables
\tabletypesize{\tiny}

\definecolor{bound}{rgb}{0.85,1,0.85}
\definecolor{uncertain}{rgb}{1,1,0.85}
\definecolor{unbound}{rgb}{1,.85,.85}
\renewcommand*{\arraystretch}{1.4}
\begin{deluxetable*}{cccccccccccccc}

\tablecaption{\label{tab:distances}Photometric parallax estimates of KOIs and nearby stars}
\tablehead{
\colhead{KOI} & \colhead{Sep.} & \colhead{Est. Prim.\textsuperscript{a} }& \colhead{Prim. Dist.} & \colhead{Est. Sec.\textsuperscript{a}} & \colhead{Sec. Dist.} & \colhead{$\sigma_{unassoc}$} & \colhead{$\Delta$m$_{g'}$\textsuperscript{b}} & \colhead{$\Delta$m$_{r'}$\textsuperscript{c}} & \colhead{$\Delta$m$_{i'}$\textsuperscript{c}} & \colhead{$\Delta$m$_{z'}$\textsuperscript{c}} & \colhead{$\Delta$m$_{J}$\textsuperscript{d}} & \colhead{$\Delta$m$_{H}$\textsuperscript{d}} & \colhead{$\Delta$m$_{K}$\textsuperscript{d}} \\
 & \colhead{('')} & \colhead{SpT} & \colhead{(pc)} & \colhead{SpT} & \colhead{(pc)} & \colhead{} & \colhead{} & \colhead{} & \colhead{} & \colhead{} & \colhead{} & \colhead{} & \colhead{}
}
\startdata
\rowcolor{bound}
1 & 1.13 & G2 & $202^{+34}_{-41}$ & K7 & $275^{+63}_{-41}$ & 1.62 &  & 3.98$\pm$0.04 & 3.77$\pm$0.03 & 3.59$\pm$0.03 & 2.8$\pm$0.1 & 2.5$\pm$0.1 & 2.36$\pm$0.01\textsuperscript{f} \\ \rowcolor{unbound}
4 & 3.42 & F3 & $349^{+59}_{-71}$ & F5 & $1999^{+393}_{-344}$ & 7.61 &  & 4.71$\pm$0.01 & 4.61$\pm$0.03 & 4.43$\pm$0.02 & 4.23$\pm$0.01 &  &  \\ 
\rowcolor{unbound}
13 & 1.16 & B8 & $786^{+132}_{-159}$ & A4 & $414^{+90}_{-65}$ & 3.33 &  & 0.24$\pm$0.01 & 0.29$\pm$0.05 &  & 0.18$\pm$0.03 &  & 0.14$\pm$0.01\textsuperscript{f} \\
\rowcolor{bound}
42 & 1.74 & F9 & $95^{+16}_{-19}$ & K2 & $137^{+30}_{-22}$ & 1.92 &  & 2.86$\pm$0.03 & 2.68$\pm$0.02 & 2.54$\pm$0.02 & 2.21$\pm$0.03 &  & 1.87$\pm$0.01\textsuperscript{f} \\
\rowcolor{unbound}
70 & 3.86 & G6 & $272^{+46}_{-55}$ & K7 & $1086^{+171}_{-217}$ & 6.68 &  & 5.75$\pm$0.04 & 5.2$\pm$0.04 & 4.54$\pm$0.12 & 4.44$\pm$0.02 &  & 4.14$\pm$0.03\textsuperscript{f} \\ 
\rowcolor{unbound}
75 & 3.47 & G0 & $162^{+27}_{-33}$ & G2 & $3224^{+577}_{-594}$ & 9.94 &  & 6.92$\pm$0.23 & 6.62$\pm$0.12 & 6.58$\pm$0.24 & 6.65$\pm$0.1 &  & 6.51$\pm$0.04\textsuperscript{f} \\ 
\rowcolor{unbound}
97 & 1.9 & F5 & $573^{+96}_{-116}$ & F5 & $3712^{+726}_{-642}$ & 7.95 &  & 4.36$\pm$0.03 & 4.27$\pm$0.05 & 4.26$\pm$0.07 &  &  &  \\ 
\rowcolor{unbound}
155 & 4.01 & G1 & $516^{+87}_{-104}$ & K4 & $1227^{+155}_{-271}$ & 4.61 & 4.33$\pm$0.05 & 4.38$\pm$0.17 & 4.04$\pm$0.19 & 3.76$\pm$0.19 &  &  &  \\
\enddata
\begin{small}\tablecomments{Rows are colored by the probability of physical association of each nearby star to the KOI: green are likely bound, red are likely unbound, and orange are uncertain.
\tablenotetext{a}{Methodology described in Section$~\ref{sec:probassociation}$.}
\tablenotetext{b}{From PANSTARRs \citep{panstarrs}.}
\tablenotetext{c}{From observations at Robo-AO KP described in Section \ref{sec:observations}.}
\tablenotetext{d}{From UKIRT \citep{lawrence07} unless noted.}
\tablenotetext{e}{From California \textit{Kepler} Survey \citep{cks}.}
\tablenotetext{f}{From \citet{kraus16}.}
\tablenotetext{g}{From \citet{atkinson16}.}
\tablenotetext{h}{From Keck NIRC-2 observations, as described in Section \ref{sec:keckao}.}
\tablenotetext{i}{Blended Robo-AO photometry of close binary resolved with Keck-AO, not used in stellar characterization.}
}\end{small}
\end{deluxetable*}
\clearpage

\LongTables
\clearpage
\tabletypesize{\scriptsize}

\begin{deluxetable*}{ccccccccc}
\tablewidth{0pt}
\tablecaption{\label{tab:radii}Implications on Derived Radius of \textit{Kepler} Planetary Candidates}
\tablehead{
\colhead{Object} & \colhead{Sep.} & \colhead{$\Delta$m$_{i'}$\textsuperscript{a}} & \colhead{R$_{p}$/R$_{\star}$\textsuperscript{b}} & \colhead{R$_{\star, target}$} & \colhead{R$_{\star, secondary}$} & \colhead{R$_{p,0}$\textsuperscript{c}} & \colhead{R$_{p,prim.}$\textsuperscript{d}} & \colhead{R$_{p,sec.}$\textsuperscript{e}}\\
\colhead{} & \colhead{(\arcsec)} & \colhead{(mags)} & \colhead{} & \colhead{(R$_{\odot}$)} & \colhead{(R$_{\odot}$)} & \colhead{(R$_{\Earth}$)} & \colhead{(R$_{\Earth}$)} & \colhead{(R$_{\Earth}$)}
}
\startdata
0001.01 & 1.13 & 3.77 & 0.1239 & $1.12^{+0.12}_{-0.09}$ & $0.75^{+0.03}_{-0.03}$ & $13.0^{+0.5}_{-0.5}$ & $13.2^{+0.5}_{-0.5}$ & $49.9^{+6.3}_{-6.4}$  \\ 
0004.01 & 3.42 & 4.61 & 0.0396 & $1.39^{+0.15}_{-0.09}$ & $1.35^{+0.15}_{-0.11}$ & $12.9^{+1.8}_{-4.2}$ & $13.0^{+1.8}_{-4.2}$ & $105^{+19}_{-18}$ \\ 
0013.01 & 1.16 & 0.29 & 0.0647 & $2.48^{+0.29}_{-0.39}$ & $1.78^{+0.08}_{-0.11}$ & $21.4^{+6.3}_{-9.5}$ & $28.5^{+8.5}_{-12.6}$ & $23.3^{+5.6}_{-3.8}$  \\ 
0042.01 & 1.74 & 2.68 & 0.0166 & $1.17^{+0.11}_{-0.03}$ & $0.89^{+0.09}_{-0.03}$ & $2.43^{+0.12}_{-0.14}$ & $2.53^{+0.13}_{-0.15}$ & $6.6^{+0.9}_{-0.7}$  \\ 
0070.04 & 3.86 & 5.2 & 0.0083 & $1.04^{+0.04}_{-0.08}$ & $0.75^{+0.04}_{-0.07}$ & $0.84^{+0.09}_{-0.08}$ & $0.84^{+0.09}_{-0.08}$ & $6.7^{+0.9}_{-0.9}$  \\ 
0070.05 & 3.86 & 5.2 & 0.0089 & $1.04^{+0.04}_{-0.08}$ & $0.75^{+0.04}_{-0.08}$ & $0.9^{+0.1}_{-0.08}$ & $0.9^{+0.1}_{-0.08}$ & $7.1^{+1.0}_{-1.0}$  \\ 
0070.01 & 3.86 & 5.2 & 0.0296 & $1.04^{+0.04}_{-0.08}$ & $0.75^{+0.03}_{-0.09}$ & $2.99^{+0.33}_{-0.27}$ & $3.0^{+0.33}_{-0.27}$ & $23.7^{+2.8}_{-3.6}$  \\ 
0070.02 & 3.86 & 5.2 & 0.0176 & $1.04^{+0.04}_{-0.08}$ & $0.75^{+0.04}_{-0.07}$ & $1.78^{+0.19}_{-0.16}$ & $1.79^{+0.19}_{-0.16}$ & $14.1^{+2.0}_{-1.9}$  \\
\enddata
\tablecomments{
\tablenotetext{a}{From observations at Robo-AO KP described in Section \ref{sec:observations}.}
\tablenotetext{b}{Diluted ratio of planet radius to stellar radius, from NASA Exoplanet Archive.}
\tablenotetext{c}{Original planetary radius estimate, from NASA Exoplanet Archive.}
\tablenotetext{d}{Estimated eclipsing object radius in the scenario where it is physically bound to the target star, corrected for transit dilution caused by the presence of nearby stars.}
\tablenotetext{e}{Estimated eclipsing object radius in the scenario where it is bound to the companion star, correcting for transit dilution by nearby stars and using stellar radius estimates derived from estimated spectral types in Table$~\ref{tab:distances}$ and \citet{habets81}.}
}
\end{deluxetable*}

\end{document}